\documentclass[a4paper]{jpconf}
\usepackage{graphicx}
  
\def\lsi{\raise0.3ex\hbox{$<$\kern-0.75em\raise-1.1ex\hbox{$\sim$}}}
\def\gsi{\raise0.3ex\hbox{$>$\kern-0.75em\raise-1.1ex\hbox{$\sim$}}}

\newcommand{\gsim}{\mathop{\gsi}}

\begin{document}

\title{If $\theta_{13}$ is large, then what?
}

\author{Hisakazu Minakata}

\address{Department of Physics, Tokyo Metropolitan University, 
Hachioji, Tokyo 192-0397, Japan}

\ead{hisakazu.minakata@gmail.com}

\begin{abstract}
As indicated by the recent data obtained by the T2K and the MINOS experiments 
$\theta_{13}$ can be large, even be comparable to the Chooz limit. 
Assuming that it will be confirmed by ongoing reactor and accelerator experiments  
I discuss its possible implications in the context of how to explore the remaining 
unknowns in the lepton flavor mixing. 
While it opens wide range of possibilities to explore CP and the mass hierarchy, 
I emphasize that the large $\theta_{13}$ allows us to take ``all in one'' 
(everything in a lunch box) approach. 
\end{abstract}

\section{It was a speculative title when it was proposed}

You might have suspected that my title is too timely, but as a matter of fact, it was 
a speculative one when it was proposed to one of the conveners in early May, much 
earlier than the announcement of $\nu_e$ appearance events at T2K \cite{T2K-indication}. 
The resulted six clean events which fulfill the criteria for electron-like events in SK 
strongly suggest that $\theta_{13}$ has a nonzero value. Their best fit value is 
around $\theta_{13} \sim 10^{\circ}$ close to the Chooz limit \cite{CHOOZ}. 
It was soon followed by the MINOS collaboration which also obtained a relatively 
large value of $\theta_{13}$ \cite{MINOS-indication}. 
In fact, by taking account of these new data from T2K and MINOS into the global analysis 
the Bari group concluded that it implies evidence for nonzero $\theta_{13}$ at confidence 
level (CL) higher than $3 \sigma$ \cite{bari-global}. 
The best fit value is $\theta_{13} \simeq 8^{\circ}$. See also \cite{valencia-update}.
I believe it an urgent matter to discuss what are the consequences of 
suggested large values of $\theta_{13}$ though it must be confirmed by 
the further running of these experiments as well as the ongoing reactor $\theta_{13}$ 
experiments \cite{DCHOOZ,RENO,Daya-Bay}.

\section {Why $\theta_{13}$ so large?}

I stated in a talk in Neutrino 2008 in Christchurch that 
``my personal prejudice is that $\theta_{13}$ is relatively large. 
If it is true it further encourages us to proceed to search for lepton 
CP violation and determination of the mass hierarchy'' \cite{nu2008-mina}.
Here is a brief recapitulation of the very simple reasoning which led me to the 
above statement. 
The lepton flavor mixing matrix, the MNS matrix \cite{MNS}, 
$U_{MNS}$, is a ``relative matrix'' $U_{\ell}^{\dagger} U_{\nu}$, where $U_{\ell}$ and 
$U_{\nu}$ denote the unitary matrices which diagonalize the charged lepton and 
the neutrino mass matrices, respectively. 
The MNS matrix is composed of the three angles $\theta_{ij}$ ($i, j = 1-3$) and a 
CP violating phase $\delta_{\ell}$, the lepton analogue of the quark 
Kobayashi-Maskawa phase $\delta_{q}$ \cite{KM}.
Two of the angles are known to be large, $\theta_{12} \simeq 34^{\circ}$ and 
$\theta_{23} \simeq 45^{\circ}$. Then, it is highly unlikely that only the remaining 
angle $\theta_{13}$ is extremely small, unless there is a good reason for tuning only 
$\theta_{13}$ into a small value. 
That is, if a symmetry exists such that $\theta_{13}$ vanishes at the symmetry limit, 
there is a natural reason why only $\theta_{13}$ is small among the three mixing angles. 
There exists many examples for such symmetry including $\mu - \tau$ symmetry and 
discrete symmetries. For a review see, e.g. \cite{moha-smi}. 
It is my prejudice that the large $\theta_{13}$ suggested by T2K and MINOS 
measurement, if confirmed, would give a tough time for these symmetries which 
aims at explaining small $\theta_{13}$, unless a large correction to the symmetry limit 
is shown to be naturally induced.

\section{Wide range of possibilities are now open for measuring CP and mass hierarchy}

Without having compelling reasons for large $\theta_{13}$, one had to take into account 
the possibility that $\theta_{13}$ is small in thinking about ways for measurement of 
CP phase and determination of the neutrino mass hierarchy. 
That is why many of the existing proposals for such apparatus naturally include those 
that have sensitivities to a tiny value of $\theta_{13}$ with varying limiting values. 
They include, for example (sorry for my arbitrary choice), neutrino factory \cite{nufact,golden}, 
beta beam \cite{beta}, superbeam enriched with two detectors \cite{T2KK}, 
and their combination \cite{MEMPHYS}. For their sensitivities see e.g.,  \cite{ISS-nufact}. 
I wonder, however, that racing for $\theta_{13}$ sensitivity could have produced an 
atmosphere that the better facility is the one which has sensitivities to smaller values 
of $\theta_{13}$. In seeing the T2K and MINOS indication for large $\theta_{13}$, 
I believe that it is the time to start thinking about changing the gear to more relevant things.

What is good if $\theta_{13}$ is as large as $\sim 10^{\circ}$? 
My first answer is that a wide range of possibilities becomes open for strategies and 
setups aiming at exploring lepton CP violation through neutrino oscillation. 
I believe that in the context of conventional superbeam \cite{superbeam} 
unambiguous evidence for lepton CP violation requires a megaton scale water 
Cherenkov detector, or a $\sim 100$ kton liquid Ar detector \cite{T2K-LOI,LBNE,liquidAr}. 
Yet, the suggested large value of $\theta_{13}$, still need to be confirmed and 
assumed here, stimulates to examine other possibilities such as 
reactor-accelerator combined method \cite{reactor-CP04}, 
combining different accelerator measurement \cite{MNP3,BMW03,combining-expts}, 
or the one adding one more new facility which may or may not require an 
extensive cost \cite{MEMPHYS,combining-channels}. 
Some more exotic options are discussed in \cite{nu2008-mina}. 
It is notable that the CERN-MEMPHYS project, despite an almost vacuum setting, has a 
 reasonable sensitivity to the mass hierarchy to $\sin^2 2\theta_{13} \simeq 0.03$
before combining with atmospheric neutrino data \cite{MEMPHYS-phys-pot}.

Sometime ago the CP sensitivity achievable by the reactor-accelerator combined 
method was examined in detail \cite{1st-hint}. The results obtained in this analysis 
indicate that combining data taken and to be taken by all planned experiments 
is not enough to guarantee discovery of CP violation in $\delta_{\ell}$ coverage of 
more than 30\% in most of the allowed region of $\theta_{13}$ even though one 
assumes that the mass hierarchy is known. 
It should be noticed, as observed and emphasized in \cite{reactor-CP04}, that 
the CP sensitivity is severely damaged by the parameter degeneracy due to 
unknown neutrino mass hierarchies \cite{MNjhep01}. 
For a recent overview of the parameter degeneracy see \cite{MU-Pdege}.

\section{``All in one'' approach becomes feasible}

I believe that the one of the most important implications of the assumed large 
value of $\theta_{13}$ is that one can think of ``all in one'' 
({\em obento} in Japanese, everything in a lunch box) approach. 
As we saw in the above, a traditional approaches for CP and the mass hierarchy 
entailed in aggressive proposals \cite{nufact,beta,T2KK,MEMPHYS} 
to include a very small value of $\theta_{13}$ within the scope. 
Though powerful and the redundancy is of course welcome, they may not be the 
optimal cost-effective apparatus for large $\theta_{13}$. 
To my opinion, what is good in living in the world with large $\theta_{13}$ is that 
one can achieve the same goal by less expensive ``all in one'' setting. 
More concretely, I mean by it a single detector 
(such as a megaton water Cherenkov detector) assuming prior existence of 
intense neutrino beam. 
Timely enough the LOI of Hyper-Kamiokande (Hyper-K) project has just 
appeared \cite{HK-LOI}. Therefore, let us briefly discuss the spirit of the ``all in one'' 
approach by taking this concrete setting.

With intense neutrino beam from 1.66 MW proton driver at J-PARC it is demonstrated 
that Hyper-K with 0.56 megaton fiducial mass has a superb performance for 
discovery of CP violation, which covers most of the region of $\theta_{13}$ 
allowed at 3$\sigma$ CL \cite{bari-global}.
See Fig.~24 of \cite{HK-LOI} which assumes the known mass hierarchy and for 
1.5 (3.5) years of neutrino (antineutrino) running. 
It shows a remarkable accuracy of determination of $\delta_{\ell}$ with 1$\sigma$ 
error smaller than 20 degree for $\sin^2 2\theta_{13} \gsim 0.03$. 
However, if the mass hierarchy is unknown the analysis suffers from the 
sign-$\Delta m^2$ degeneracy \cite{MNjhep01}.

While allowing clean measurement of CP violation, a drawback of setting with such 
short baseline is that 
it is difficult to determine the mass hierarchy, another important goal of future neutrino experiments. That is why many proposals for determining all the unknowns in neutrino parameters exploit baselines of several thousand km. If $\theta_{13}$ is large, however, 
the mass hierarchy can be determined by high-statistics observation of atmospheric neutrinos, the idea discussed by many authors. 
See e.g., \cite{peres-smirnov03,bernabeu03} for early references. 
The analysis presented in \cite{HK-LOI} indicates that the sensitivity to the mass hierarchy depend very much on which octant $\theta_{23}$ lives but it can be carried out 
by $\sim 3-10$ years running of Hyper-K  for large $\theta_{13}$. 
It should also be mentioned that having other capabilities such as proton decay 
discovery is of crucial importance. 
I emphasize that the realization of ``all in one'' approach is, of course, not limited to 
this particular version discussed above. But, it is certainly encouraging to see the 
concrete example appeared which is armed with the realistic simulations based on 
their long-term experiences at Super-K.

\section{Large-$\theta_{13}$ perturbation theory of neutrino oscillation}

Is theoretical treatment of neutrino oscillation well oiled when a large value of 
$\theta_{13}$ close to the Chooz limit is established? 
For example, does the well known treatment of the degeneracy \cite{MU-Pdege} 
need modification? If yes, to what extent?

\begin{figure}
\centerline{\includegraphics[width=6.4cm]{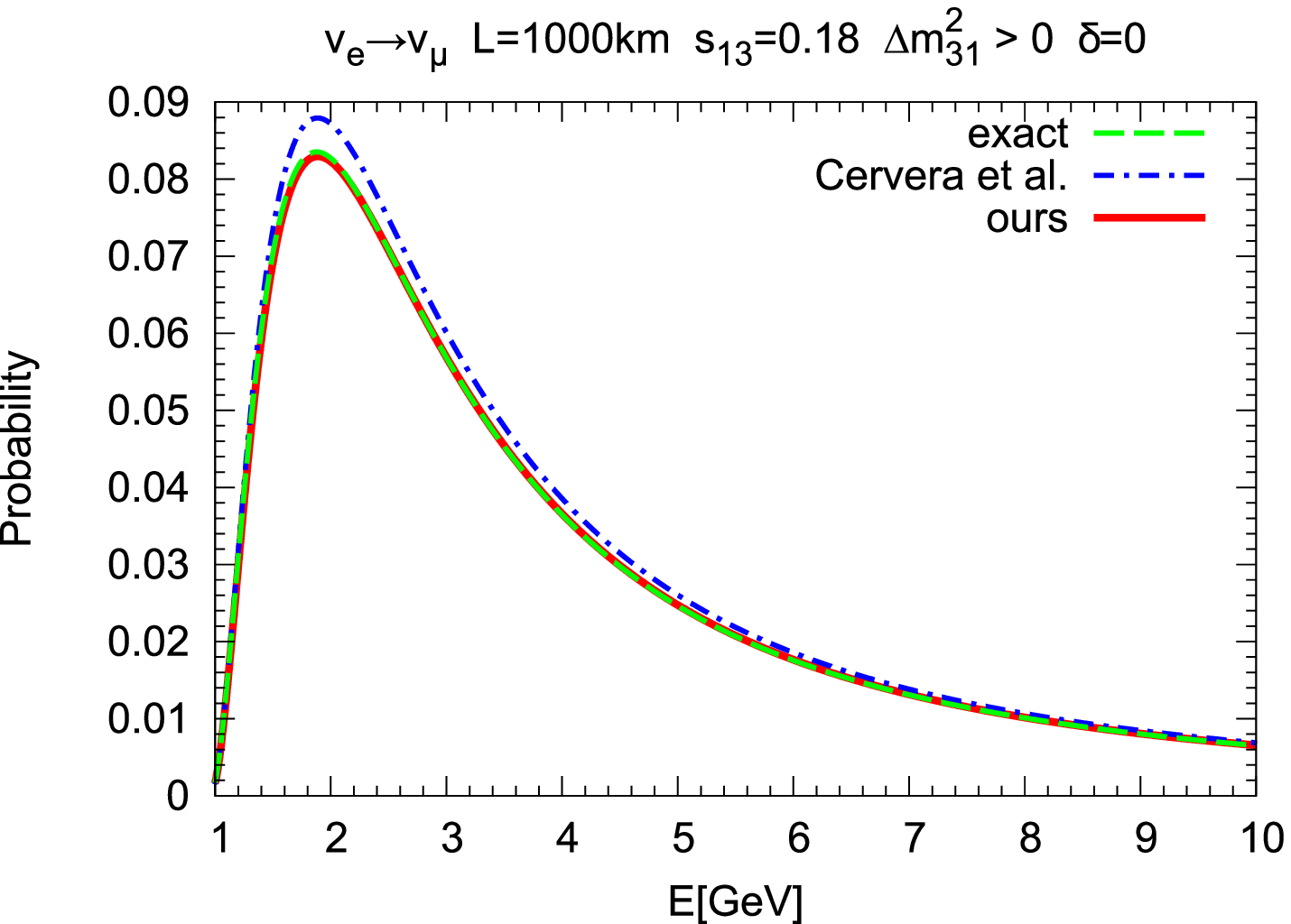} \includegraphics[width=6.4cm]{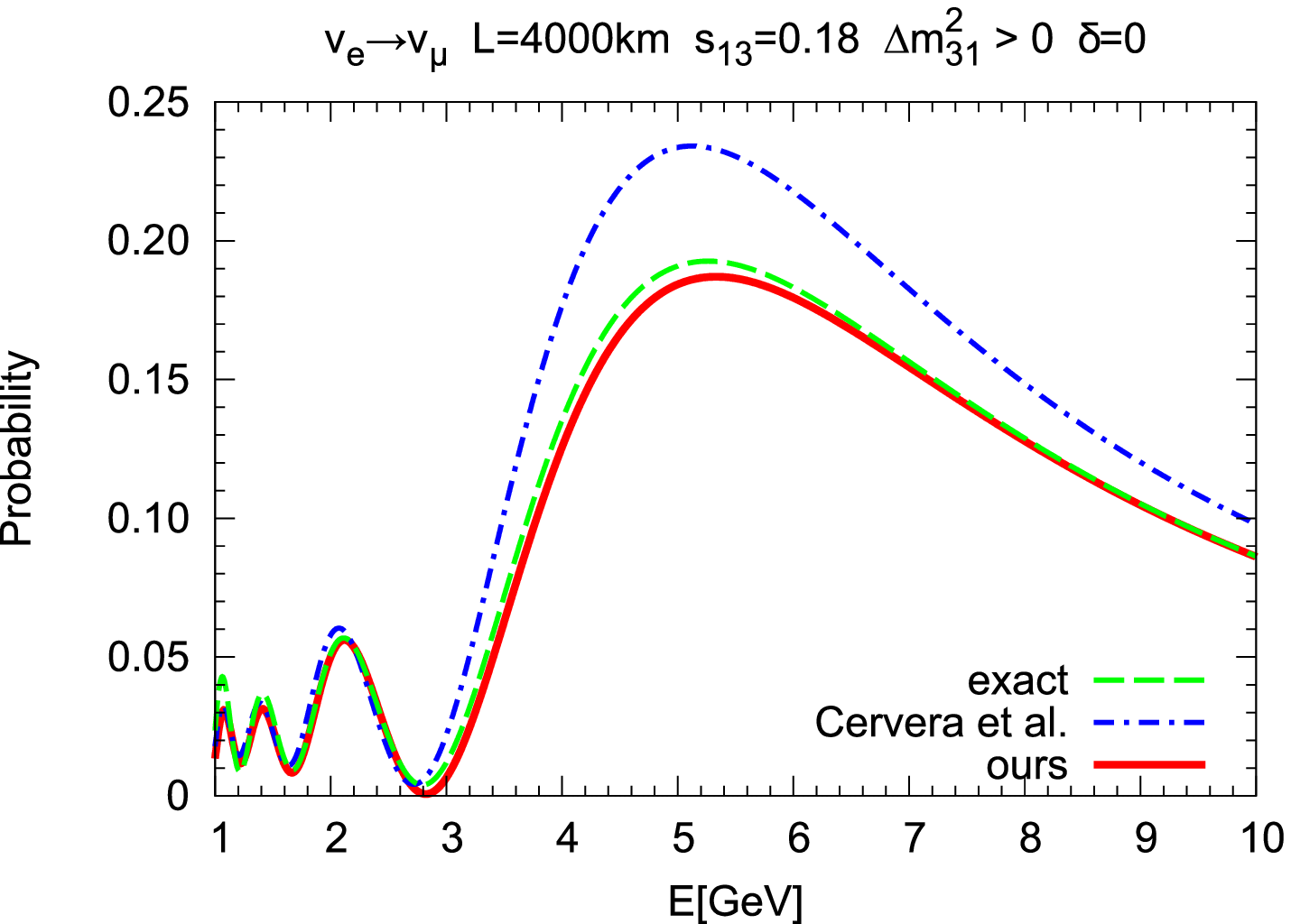} }
\caption{Comparison between the exact oscillation probability $P (\nu_{e} \rightarrow \nu_{\mu})$ 
computed numerically as a function of energy (green dashed line), 
the one calculated by the Cervera {\it et al.} formula (blue dash-dotted line), 
and with our formula with large $\theta_{13}$ corrections (red solid line). 
The left and the right panels are for baselines $L=1000$ km and $L=4000$ km 
for which the matter density is taken as 2.8 g/cm$^3$ and 3.6 g/cm$^3$, respectively. 
} 
\label{fPemu_1000-4000}
\end{figure}

To answer these questions a framework dubbed as ``$\sqrt{\epsilon}$ perturbation theory'' 
($\epsilon \equiv \Delta m^2_{21} / \Delta m^2_{31}$)
has been formulated \cite{large-theta13-perturbation} by assuming 
$s_{13} \sim \sqrt{\epsilon} \simeq 0.18$, which roughly corresponds to the Chooz limit. 
By doing so one can systematically compute the large-$\theta_{13}$ corrections to the 
Cervera {\it et al.} formula \cite{golden}. 
While large corrections arise in certain limited region of energy and baseline as 
indicated in Fig.~\ref{fPemu_1000-4000}, one can show on general ground that the correction terms 
are of order $\sim \epsilon^2 \simeq 10^{-3}$.
(By general argument one can show that they are the terms proportional to either 
$s^4_{13}$ or $\epsilon s^2_{13}$  \cite{large-theta13-perturbation}.)
Therefore, the large-$\theta_{13}$ correction to the degenerate solutions obtained 
with the Cervera {\it et al.} formula, generally speaking, is not sizable. 
See \cite{large-theta13-perturbation} for more detail.

\section{Conclusion}

Assuming that the T2K and MINOS hints for large $\theta_{13}$ will be confirmed, 
I argued that we are at the turning point of thinking of our strategy of 
how to explore the remaining unknowns in the lepton flavor mixing. 
In particular, it allows us to rely on ``all in one'' approach. 
I hope that the personal view can be strengthened by further exploration toward Nufact 2012.

\ack
This work was supported in part by KAKENHI, Grant-in-Aid for
  Scientific Research No. 23540315, Japan Society for the Promotion of Science.


\section*{References}

\vspace{0.5cm}

\end{document}